\documentclass[11pt,twoside]{article}
\usepackage[pdftex]{graphicx}
\usepackage{amsmath}
\usepackage{amssymb}
\usepackage{cite}
\begin{document}

\newcommand{\pst}{\hspace*{1.5em}}

\newcommand{\be}{\begin{equation}}
\newcommand{\ee}{\end{equation}}
\newcommand{\bm}{\boldmath}
\newcommand{\ds}{\displaystyle}
\newcommand{\bea}{\begin{eqnarray}}
\newcommand{\eea}{\end{eqnarray}}
\newcommand{\ba}{\begin{array}}
\newcommand{\ea}{\end{array}}
\newcommand{\arcsinh}{\mathop{\rm arcsinh}\nolimits}
\newcommand{\arctanh}{\mathop{\rm arctanh}\nolimits}
\newcommand{\bc}{\begin{center}}
\newcommand{\ec}{\end{center}}

\thispagestyle{plain}

\label{sh}


\begin{center} {\Large \bf
\begin{tabular}{c}
QUANTUM CORRELATIONS FOR TWO COUPLED 
 \\[-1mm]
OSCILLATORS INTERACTING WITH TWO HEAT BATHS
\end{tabular}
 } \end{center}

\bigskip

\bigskip

\begin{center} {\bf
Ivan V. Dudinets$^{1*}$ and Vladimir I. Man'ko $^{1,2}$
}\end{center}

\begin{center}
{\it
$^1$Moscow Institute of Physics and Technology (State University)\\
Institutskii per. 9, Dolgoprudnii, Moscow Region 141700, Russia

\smallskip

$^2$P.N. Lebedev Physical Institute, Russian Academy of Sciences\\
Leninskii Prospect 53, Moscow 119991, Russia
}
\smallskip

$^*$Corresponding author e-mail: dudinets@phystech.edu \\
\end{center}

\begin{abstract}\noindent
We study a system of two coupled oscillators (the $A$ oscillator) each of the oscillators linearly interacts with its own heat bath consisting of a set of independent harmonic oscillators (the $B$ oscillators). The initial state of the $A$ oscillator is taken to be coherent while the $B$ oscillators are in thermal states. We analyze the time-dependent state of the $A$ oscillator which is a two-mode Gaussian state. By making use of Simon's separability criterion we show that this state is separable for all times. We consider the equilibrium state of the $A$ oscillator in detail and calculate its Wigner function. 

\end{abstract}

\medskip

\section{Introduction}
\pst

Entangled continuous-variable states are of great importance in quantum informational theory due to its various practical applications (see, e.g.,~\cite{braunstein2005quantum}). In this regard, entangled Gaussian states are of great importance as well. Moreover, Gaussian states, especially two-mode Gaussian states, are the most studied ones among continuous-variable states in the sense that there exist separability  criterion~\cite{simon2000peres} and explicit formulas of different measures of entanglement (see, e.g.,~\cite{adesso2007entanglement,plenio2005introduction}) for two-mode Gaussian states. However, in real experiments the interaction of the system under study with the environment is unavoidable. This interaction leads to the degradation of the entanglement in the system~\cite{li2005generation}, i.e. the amount of entanglement decreases. In these papers the initial state of the system under consideration is assumed to be entangled. On the other hand, initially separable states can become entangled in the presence of dissipation. This phenomenon called the emergence of entanglement is a subject of various studies (see, e.g.,~\cite{de2008effect}). Thus preservation of entanglement in dissipative systems received much attention. There exist different approaches to study systems with dissipation, for example, an approach based on the quantization through the classical constant of motion  was suggested in~\cite{lopez2004quantum}. Another approach is based on the interaction of the system in question with a heat bath. One of the frequently used model of a heat bath is a collection of harmonic oscillators~\cite{glauber1984damping}.	
The concept of squeezed states to study the dynamics of weakly coupled oscillators was applied in~\cite{alekseev2009squeezed}. 
The model of an oscillator interacting with  a heat reservoir, which also
consists of oscillators was the subject of investigation in various papers~\cite{dodonov1995quantum, tatarskiui1987example}.
The path integration method was used in~\cite{dorofeyev2014quasi,dorofeyev2016dynamics,dorofeyev2014relaxation} to analyze the evolution of quantum oscillators interacting with separate reservoirs. 
The dynamics of a qubit coupled to a large number of harmonic oscillators was studied in~\cite{chakravarty1984dynamics}. 
The entanglement evolution of two coupled quantum oscillators interacting with one reservoir or two reservoirs analyzed in~\cite{liu2007non,freitas2012dynamics,de2008effect}. 
The dynamics of networks of oscillators connected to the external environment modeled by a set of independent oscillators was considered in~\cite{martinez2013dynamics,glauber1984damping}. 

In this paper we study two interacting quantum harmonic oscillators (the $A$ oscillator) each coupled to its own heat bath characterized by its own temperature. Each of the baths is modeled by an infinite set of independent harmonic oscillators (the B oscillators). We assume that the interaction between the $A$ oscillator and the $B$ oscillators is linear in their position and momentum operators. The evolution of the system is such that the state of the $A$ oscillator being Guassian at initial time remains being Gaussian for all times. For long times the $A$ oscillator reaches the equilibrium state which is not Gibbsian and characterized by the temperatures of the baths. The aim of the present paper is the analysis of the entanglement properties of the equilibrium state reached by the A oscillator. 

The paper is organised as follows. In Sec~2, following~\cite{glauber1984damping}, we study the evolution of the density operator of the $A$ oscillator which can be obtained by reducing the density operator of the entire system over the variables of heat baths. For special choice of the initial state of the system the state obtained has a Gaussian form and thus fully characterized by its covariance matrix. We give an explicit expression for the time-dependent covariance matrix of the $A$ oscillator. Using Simon's separability criterion we prove that the state of the $A$ oscillator is separable for all times. In Sec.~3 we show that the $A$ oscillator for long times reaches the equilibrium state and calculate the covariance matrix of this state. In Sec.~4 we obtain the Wigner function for this equilibrium state. Then we consider the equilibrium state for the case of equal temperatures of the baths. Finally, a brief summary is given in Sec.~5.

\section{The density operator of the two coupled oscillators}
\pst
Following~\cite{glauber1984damping}, let us consider a system of two coupled quantum oscillators $A_1$ and $A_2$. We designate the two oscillators in question as the $A$ oscillator. Each of the two oscillators linearly interacts with its own heat bath each of which modeled by the infinite set of oscillators denoted below as the $B_1$ and $B_2$ oscillators. The temperatures of the baths are different in general. We denote the oscillators of a heat bath containing a system of the two aforehead mentioned baths as the $B$ oscillators. 
 We take the Hamiltonian of the system under consideration to be of the form

\begin{eqnarray}\label{Hamilt}
\hat{H}=\omega\left(\hat{a}^{\dagger}_1 \hat{a}_1+\hat{a}^{\dagger}_2 \hat{a}_2 \right )+\lambda\left(\hat{a}^{\dagger}_1 \hat{a}_2+\hat{a}^{\dagger}_2 \hat{a}_1 \right )+\sum_{k}\omega_k\left(\hat{b}^{\dagger}_{1k} \hat{b}_{1k}+\hat{b}^{\dagger}_{2k} \hat{b}_{2k} \right )\nonumber\\
+\sum_{k}\left[\lambda_k\left(\hat{a}^{\dagger}_{1} \hat{b}_{1k}+\hat{a}^{\dagger}_{2k} \hat{b}_{2k} \right )+
\lambda^*_k\left(\hat{a}_{1} \hat{b}^{\dagger}_{1k}+\hat{a}_{2k} \hat{b}_{2k}^{\dagger} \right )\right],
\end{eqnarray}
where $\hat{a}^{\dagger}_i$, $\hat{a}_i$, $\hat{b}^{\dagger}_{ik}$ and $\hat{b}_{ik}$ are the creation and annihilation operators of the $A_i$ and $B_i$ oscillators, respectively, $\omega$  and $\omega_k$ are the frequencies of the $A$ oscillator and $B$ oscillators, $\lambda$ is a coupling constant of the $A_1$ and $A_2$ oscillators, $\lambda_k$ are a set of complex coupling constants, and $\lambda^*_k $ are the complex conjugates of $\lambda_k$.
We assume that the interaction of each oscillator $A_i$ with its heat bath is symmetric and the parameters $\omega_k$ and $\lambda_k$ are the same for each $A_i$ oscillator.
 Here and throughout the paper the subscript $i$ takes two values, $1$ and $2$, the subscript $k$ labels oscillators in the $i$th heat bath and vary from $1$ to $\infty$. The subscript $k$ can be continuous, in this case the summation over $k$ is assumed to be an integration over the continuous $k$. The Planck and the Boltzmann constants are taken to be unity, i.e. $\hbar=1$ and $k_B=1$. It is necessary to point out that the Hamiltonian is written in the rotated-wave approximation form in which the effect of the antiresonance terms $\hat{a}^{\dagger}_{i} \hat{b}^{\dagger}_{ik}$ and $\hat{a}_{i} \hat{b}_{ik}$, $i=1,2$ is neglected. 

Our purpose is to obtain, following~\cite{glauber1984damping}, the time-dependent density operator of the $A$ oscillator. To do that, let us first specify its initial state. We assume that the $A$ oscillator initially in the pure coherent state $|\alpha_1, \alpha_2\rangle=|\alpha_1\rangle |\alpha_2\rangle$ (that is, $|\alpha_i\rangle$ is an eigenstate of the annihilation operator  $\hat{a}_i $ with the complex eigenvalue $\alpha_i$, i.e. $\hat{a}_i |\alpha_i\rangle = \alpha_i |\alpha_i\rangle$), whereas the $B$ oscillators are initially in thermal states. The reduced density operator $\hat{\rho}_{ik}$ of the $k$th oscillator in the $B_i$ heat bath is given by
\be
\hat{\rho}_{ik}=\int \exp{\left(-\frac{|\beta_{ik}|^2}{\langle n_{ik}\rangle}\right)}|\beta_{ik}\rangle \langle \beta_{ik}|\frac{d^2\beta_{ik}}{\pi\langle n_{ik}\rangle},
\ee
in which  $\langle n_{ik}\rangle$ is the mean number of quanta. Here $d^2\beta_{ik}$ means that the integration is performed over the real and imaginary parts of the complex number $\beta_{ik}$, $d^2\beta_{ik}=d (\textit{Re} \,\beta_{ik})\,d(\textit{Im}\, \beta_{ik})$. If the $B_i$ heat bath is characterized by a temperature $T_i$, then the mean number of quanta is described by the Plank distribution
\be
\langle n_{ik}\rangle=\left( \exp\left(\frac{\omega_k}{T_i}\right)-1 \right)^{-1}.
\ee
The density operator of the entire system comprising the $A$ and $B$ oscillators at initial time $t=0$ is assumed to have the factorized form 
\begin{eqnarray}\label{initial_density_matrix}
\hat{\rho}(0)=|\alpha_1,\alpha_2\rangle \langle \alpha_1,\alpha_2|\prod _{ik} \hat{\rho}_{ik}\nonumber\\
=\int |\alpha_1, \alpha_2, \{\beta_{ik}\}\rangle \langle\alpha_1, \alpha_2, \{\beta_{ik}\}|
\prod_{ik}\exp\left(-\frac{|\beta_{ik}|^2}{\langle n_{ik}\rangle}\right)\frac{d^2\beta_{ik}}{\pi\langle n_{ik}\rangle}
\end{eqnarray}
Here $|\alpha_1, \alpha_2, \{\beta_{ik}\}\rangle $ denotes the direct product of coherent states 
\be
|\alpha_1, \alpha_2, \{\beta_{ik}\}\rangle=|\alpha_1\rangle|\alpha_2\rangle\prod_{ik}|\beta_{ik}\rangle.
\ee

The evolution of the density operator $\hat{\rho}(t)$ of the entire system is governed by the von-Neumann equation 
\be\label{Neumann}
i\frac{\partial \hat{\rho}(t)}{\partial t}=\left[ \hat{H},\hat{\rho}(t)\right]
\ee
with an initial condition~(\ref{initial_density_matrix}). The explicit form of the operator $\hat{\rho}(t)$ is not of interest to us, since our purpose is the calculation of the density operator for the $A$ oscillator defined as the partial trace of $\hat{\rho}(t)$ over all the coordinates of the $B$ oscillators, namely 
\be
\hat{\rho}_A (t)=\mbox{Tr}_B \hat{\rho}(t).
\ee 
The expression for $\hat{\rho}_A (t)$ is the following~\cite{glauber1984damping}
\be\label{density_operatorA}
\hat{\rho}_A (t)=\int |\alpha_1(t),\alpha_2(t) \rangle \langle \alpha_1(t),\alpha_2(t)| \prod_{ik}\exp\left(-\frac{|\beta_{ik}|^2}{\langle n_{ik}\rangle}\right)\frac{d^2\beta_{ik}}{\pi\langle n_{ik}\rangle},
\ee
where the time-dependent functions $\alpha_1(t)$ and $\alpha_2(t)$ are given by
\begin{eqnarray}\label{alphatime}
\alpha_1(t)=\alpha_1 \left(u_1(t)+u_2(t)\right)/2+\alpha_2 \left(u_1(t)-u_2(t)\right)/2+\sum_k \beta_{1k} \left( 
v_{1k}(t)+v_{2k}(t) \right)/2+\beta_{2k} \left( v_{1k}(t)-v_{2k}(t) \right)/2\nonumber\\
\alpha_2(t)=\alpha_1 \left(u_1(t)-u_2(t)\right)/2+\alpha_2 \left(u_1(t)+u_2(t)\right)/2+\sum_k \beta_{1k} \left( 
v_{1k}(t)-v_{2k}(t) \right)/2+\beta_{2k} \left( v_{1k}(t)+v_{2k}(t) \right)/2. 
\end{eqnarray}
The explicit expression for the functions $u_i(t)$ and $v_{ik}(t)$ is given in the next section, in this section we are not interested in the form of this functions. 
It is worth noting that the state of the $A$ oscillator is Gaussian at any time $t\geq 0$ and determined by only the averages and the second moments. 

It is clear from~(\ref{density_operatorA}) that the reduced density operator $\hat{\rho}_A (t)$ does not factorize for $t\neq 0$ since the functions $\alpha_1(t)$ and $\alpha_2(t)$ depend linearly on the $\beta_{ik}$. Therefore, the question arises if the state of the $A$ oscillator is entangled or separable? In order to answer the question, it is convenient to introduce the following covariance matrix
\be\label{cov_matrix}
Q(t)= \begin{pmatrix}
  Q_{x_1x_1}& Q_{x_1p_1}& Q_{x_1x_2}& Q_{x_1p_2}\\
  Q_{x_1p_1}& Q_{p_1p_1} & Q_{x_2p_1}& Q_{p_1p_2}\\
  Q_{x_1x_2}& Q_{x_2p_1}& Q_{x_2x_2}& Q_{x_2p_2}\\
  Q_{x_1p_2}& Q_{p_1p_2} & Q_{x_2p_2}& Q_{p_2p_2}\\
 \end{pmatrix}.
 \ee
The elements of the covariance matrix are defined as $Q_{ab}=\mbox{Tr}_A\left(\hat{\rho}_A (t) \hat{a} \,\hat{b} \right)-\mbox{Tr}_A\left(\hat{\rho}_A (t)  \hat{a} \right)\mbox{Tr}_A\left(\hat{\rho}_A (t)  \hat{b} \right)$, where the trace is taken over the coordinates of the $A$ oscillator.  
Here $\hat{x}_i$ and $\hat{p}_i$ are the position and momentum operators of the $A_i$ oscillator, namely $\hat{x}_i=(\hat{a}_i+\hat{a}^\dagger_i)/\sqrt{2\omega}$, $\hat{p}_i=\sqrt{\omega}(\hat{a}_i-\hat{a}^\dagger_i)/i\sqrt{2}$. The entries of the covariance matrix for the $A$ oscillator can be easily obtained from~(\ref{density_operatorA}),~(\ref{alphatime}) and are given by
\begin{eqnarray}\label{elements_cov_matr}
\omega Q_{x_1x_1}(t)=\frac{1}{\omega} Q_{p_1p_1}(t)=\frac{1}{2}+\frac{1}{4}\sum_k \langle n_{1k}\rangle |v_{1k}(t)+v_{2k}(t)|^2+ \langle n_{2k}\rangle|v_{1k}(t)-v_{2k}(t)|^2,\nonumber\\
\omega Q_{x_2x_2}(t)=\frac{1}{\omega} Q_{p_2p_2}(t)=\frac{1}{2}+\frac{1}{4}\sum_k \langle n_{1k}\rangle |v_{1k}(t)-v_{2k}(t)|^2+ \langle n_{2k}\rangle|v_{1k}(t)+v_{2k}(t)|^2,\nonumber\\
\omega Q_{x_1x_2}(t)=\frac{1}{\omega}Q_{p_1p_2}(t)=\frac{1}{4}\sum_k \left( |v_{1k}(t)|^2-|v_{2k}(t)|^2\right)\left( \langle n_{1k}\rangle + \langle n_{2k}\rangle\right),\nonumber\\
Q_{x_1p_2}(t)=-Q_{x_2p_1}(t)=\frac{1}{4i}\sum_k \left(v_{1k}(t)v^*_{2k}(t)-v^*_{1k}(t)v_{2k}(t)\right)\left( \langle n_{1k}\rangle - \langle n_{2k}\rangle\right).
\end{eqnarray}  
The other elements of the matrix $Q$ are equal to zero. It is worth noting that all the elements of $Q$ are real.
In~\cite{simon2000peres} Simon derived a criterion for bipartite Gaussian continuous-variable states. The immediate consequence of the criterion is that a bipartite Gaussian state for which the inequality
\be
\mbox{det} \begin{pmatrix}  Q_{x_1x_2}& Q_{x_1p_2}\\
 Q_{x_2p_1}& Q_{p_1p_2}\\
 \end{pmatrix} \geq 0
\ee
holds is separable.
 The latter determinant for the $A$ oscillator reads
\be
Q_{x_1x_2}(t)Q_{p_1p_2}(t)-Q_{x_1p_2}(t)Q_{x_2p_1}(t)=\omega ^2 Q^2_{x_1x_2}(t)+Q^2_{x_1p_2}(t)\geq 0.
\ee
Therefore, the state of the $A$ oscillator is separable for all times $t\geq 0$.

\section{The equilibrium state of the two coupled oscillators}
\pst
In this section we prove that  for long times the $A$ oscillator reaches an equilibrium state, which is non-Gibbsian, and obtain the explicit expression for the covariance matrix of this state. 

The functions $\alpha_i(t)$ which appear in~(\ref{density_operatorA}) are the solution of the infinite set of linear differential equations~\cite{glauber1984damping} 

\begin{eqnarray}\label{system}
\frac{d\alpha_i(t)}{dt}=-i\omega \alpha_i(t)-i\sum_k \lambda_k \beta_{ik}(t),\nonumber \\
\frac{d\beta_{ik}(t)}{dt}=-i\omega_k \beta_{ik}(t)-i\lambda^*_k \alpha_i(t)
\end{eqnarray}
with the initial conditions 
\be 
\alpha_i(0)=\alpha_i,\quad \beta_{ik}(0)=\beta_{ik}.
\ee
The general solution of this set of equations is given by~(\ref{alphatime}) (we are not interested in the form of $ \beta_{ik}(t)$ since they do not enter the state of the $A$ oscillator) and expressed in terms of the functions $u_i(t)$ and $v_{ik}(t)$ satisfying the initial conditions 
\begin{eqnarray}
u_i(0)=1,\quad v_{ik}(0)=0. 
\end{eqnarray} 
Unless we specified the form of the functions $u_i(t)$ and $v_{ik}(t)$, the formulas for the reduced density operator $\hat{\rho}_A$~(\ref{density_operatorA}) and for the elements~(\ref{elements_cov_matr}) of the covariance matrix are have so far been regarded as exact. 
The authors of the paper~\cite{glauber1984damping} made use of the Weisskopf-Wigner approximation to solve the system~(\ref{system}). The explicit expressions for the functions $u_i(t)$ and $v_{ik}(t)$  read 
\begin{eqnarray}\label{WW}
u_i(t)=\exp\left[-\varkappa_i t-i(\Omega_i +\delta \Omega_i)t\right],\nonumber \\
v_{ik}(t)=\frac{-i\lambda_k}{\varkappa_i +i(\Omega_i +\delta \Omega_i-\omega_k)}\left\{\exp\left[-i\omega_k t\right]-\exp\left[-\varkappa_i t-i(\Omega_i +\delta \Omega_i)t\right]\right\},
\end{eqnarray}
in which $\Omega_1=\omega+\lambda$, $\Omega_2=\omega-\lambda$, the damping constants $\varkappa_i$ and the frequency shifts $\delta\Omega_i$ are defined by the relation 
\be
\delta\Omega_i-i\varkappa_i=\lim_{\epsilon \to 0}\sum_{k}\frac{|\lambda_k|^2}{\Omega_i-\omega_k+i\epsilon}.
\ee

The Weisskopf-Wigner approximation requires that the frequency spectrum $\omega_k$ should be continuous near the frequencies $\Omega_i$; the constants $ \delta \Omega _i$ and $ \varkappa_i$ should be small in comparison with the frequencies $\Omega_i$ and the coupling constants $\lambda_k$. Also we assume that the coupling constants $\lambda_k$ are much smaller than $\lambda$. This assumption means that the relaxation processes corresponding to 
the interaction between the two oscillators take place quickly in contrast to the relaxation processes corresponding to the interaction with the heat baths.

Let us first evaluate the element $Q_{x_1x_1}$~(\ref{elements_cov_matr}) of the covariance matrix $Q$. It is seen from~(\ref{WW}) that the functions $|v_{ik}(t)|$ have a sharp peak at $\omega_k=\Omega_i+\delta \Omega _i$ and decrease rapidly outside a frequency band of width $\Delta \omega_k\approx 2\varkappa_i$. Moreover, the peaks are 
frequency separated from each other, that is, the values of the function $|v_{1k}(t)|$ are small in the neighborhood of the peak of $|v_{2k}(t)|$ and vice versa. Then, since the mean number of quanta $\langle n_{ik}\rangle$ is smooth  function, we may write
\be
\omega Q_{x_1x_1}(t)=\frac{1}{2}+\frac{1}{4}\left(\langle n^{\Omega_1}_{1}\rangle+\langle n^{\Omega_1}_{2}\rangle \right)\sum_k |v_{1k}(t)|^2+\frac{1}{4}\left(\langle n^{\Omega_2}_{1}\rangle+\langle n^{\Omega_2}_{2}\rangle \right)\sum_k |v_{2k}(t)|^2, 
\ee
where we have neglected the cross-terms of the form $v^*_{1k}(t)v_{2k}(t)$. Here $\langle n^{\Omega_i}_{j}\rangle$  is the mean number of quanta at frequency $\Omega_i$ and temperature $T_j$, $j=1,2$, namely
\be
\langle n^{\Omega_i}_{j}\rangle=\left(\exp\left(\frac{\Omega_i}{T_j}\right)-1 \right)^{-1}.
\ee
 Making use of the identity
\be
|u_i(t)|^2+\sum_k |v_{ik}(t)|^2 =1,
\ee
which follows from the unitarity of the evolution operator of the entire system, we obtain
\be
\omega Q_{x_1x_1}(t)=\frac{1}{2}+\frac{1}{4}\left(\langle n^{\Omega_1}_{1}\rangle+\langle n^{\Omega_1}_{2}\rangle \right)\left(
1-|u_1(t)|^2\right)
+\frac{1}{4}\left(\langle n^{\Omega_2}_{1}\rangle+\langle n^{\Omega_2}_{2}\rangle \right)\left(
1-|u_2(t)|^2\right). 
\ee
For long time $t\gg \varkappa^{-1}_i$ we have $|u_i(t)|\ll 1$, hence
\be\label{A}
\omega Q_{x_1x_1}(\infty)\equiv A=\frac{1}{2}+\frac{1}{4}\left(\langle n^{\Omega_1}_{1}\rangle+\langle n^{\Omega_1}_{2}\rangle +\langle n^{\Omega_2}_{1}\rangle+\langle n^{\Omega_2}_{2}\rangle\right). 
\ee
One can evaluate the other elements of the covariance matrix $Q$ in the same fashion
\be\label{quasistate}
Q(\infty)= \begin{pmatrix}
  \omega ^{-1}A& 0& \omega ^{-1}B& 0\\
  0& \omega A & 0& \omega B\\
  \omega ^{-1}B& 0& \omega ^{-1}A& 0\\
  0& \omega B & 0& \omega A\\
 \end{pmatrix},
 \ee
 in which $A$ is given by~(\ref{A}), whereas $B$ reads
\be\label{B}
B=\frac{1}{4}\left(\langle n^{\Omega_1}_{1}\rangle+\langle n^{\Omega_1}_{2}\rangle -\langle n^{\Omega_2}_{1}\rangle-\langle n^{\Omega_2}_{2}\rangle\right).
\ee
The obtained state determined by the covariance matrix~(\ref{quasistate}) is the equilibrium state of the $A$ oscillator and called  the intermediate equilibrium state~\cite{glauber1984damping}. One can see that this state is a non-Gibbsian since it is determined by both the temperatures $T_1$ and $T_2$. The intermediate equilibrium state is stable in the sense that if the state of the A oscillator is excited, it comes back to the specified intermediate state. 

\section{The Wigner function for the equilibrium state of the two coupled oscillators}
\pst
In this section we obtain the density matrix of the intermediate equilibrium state in position representation and verify that for the case of equal temperatures, $T_1=T_2$, the state becomes the usual Gibbsian state. This can be easily done by calculating the corresponding  Wigner function.  

The Wigner function of a quantum state represented by the density operator $\hat{\rho}$ is defined by~\cite{wigner1932quantum} (for one mode)
\be\label{onemode}
W(q,p)=\int \frac{du}{2\pi}\exp\left(ipu\right)\langle q-\frac{u}{2}|\hat{\rho}|q+\frac{u}{2}\rangle.
\ee
Here the integration goes from $-\infty$ to $\infty$. The Wigner function is a quasi-probability function on phase-space and its marginals yield the quantum probabilities for the position and momentum, i.e.
\be
\int W(q,p)dq=\langle p|\hat{\rho}|p\rangle,	\quad \int W(q,p)dp=\langle q|\hat{\rho}|q\rangle.
\ee
The Wigner function satisfies the normalization condition 
\be
\int W(q,p)dqdp=1 
\ee 
and enables to calculate the quantum-mechanical expectation values of symmetrized product of
$n$ position and $m$ momentum operators $S(\hat{x}^n \hat{p}^m)$ in the classical-like fashion~\cite{leonhardt1997measuring}
\be
\mbox{Tr}\left[\hat{\rho}S(\hat{x}^n \hat{p}^m)\right]=\int x^n p^m W(x,p) dx dp.
\ee
However, the Wigner function can not be considered as a fair probability distribution since it may admit negative values (see, e.g.,~\cite{dudinets}). It is worth noting that the volume of the
negative part of the Wigner function can be regarded as an indicator of nonclassicality of quantum states~\cite{kenfack2004negativity}.

The generalization of the Wigner function to the case of two-mode is straightforward
\be \label{Wrho}W(x_1,p_1,x_2,p_2;t)=\int \frac{du_1du_2}{4\pi^2} \exp\left(i(p_1u_1+p_2u_2)\right)
\langle x_1-\frac{u_1}{2},x_2-\frac{u_2}{2}|\hat{\rho}(t)|x_1+\frac{u_1}{2},x_2+\frac{u_2}{2}\rangle, \ee
where we included the dependence of the density operator and the corresponding Wigner function on time. By definition, a quantum state is said to be Gaussian if its Wigner function has Gaussian form. A general form of a two-mode Gaussian Wigner function reads
\be\label{Wigner_generic}
W(x_1,p_1,x_2,p_2;t)=\frac{1}{4\pi^2\sqrt{\mbox{det}\,Q (t)}}\exp\left(-\frac{1}{2}\sum_{i,j=1}^{4} (\xi_i-\langle \xi_i(t)\rangle )Q^{-1}_{\xi_i \xi_j}(t) (\xi_j-\langle \xi_j(t)\rangle )\right).
\ee
Here $Q(t)$ is the covariance matrix~(\ref{cov_matrix}), $\xi_i$ are the arguments of the Wigner function,  $(\xi_1,\xi_2,\xi_3,\xi_4)=(x_1,p_1,x_2,p_2)$, $Q^{-1}_{\xi_i \xi_j}(t)$ denotes an element of the inverse covariance matrix, four parameters $\langle \xi_i(t)\rangle$ are the average values of momentum and position operators,
$\langle \xi_1(t)\rangle=\langle \hat{x}_1\rangle $, $\langle \xi_2(t)\rangle=\langle \hat{p}_1\rangle $, 
$\langle \xi_3(t)\rangle=\langle \hat{x}_2\rangle $ and $\langle \xi_4(t)\rangle=\langle \hat{p}_2\rangle $, where the averages are calculated through the density operator of the $A$ oscillator, for example $\langle \xi_1(t)\rangle=\mbox{Tr}\left(\hat{\rho}_{A}(t) \hat{x}_1\right) $. 

The Wigner function of the intermediate equilibrium state described by the covariance matrix~(\ref{quasistate}) reads
\begin{eqnarray}\label{Wigner_A}
W_A(x_1,p_1,x_2,p_2;\infty)=\frac{\left(e^{\Omega_1/T_1}-1\right)\left(e^{\Omega_2/T_2}-1\right)}{\pi^2\left(e^{\Omega_1/T_1}+1\right)\left(e^{\Omega_2/T_2}+1\right)} \times \nonumber\\
\exp\left(-\frac{e^{\Omega_1/T_1+\Omega_2/T_2}-1}{\left(e^{\Omega_1/T_1}+1\right)\left(e^{\Omega_2/T_2}+1\right)}\left(\omega x^2_1+\omega^{-1} p^2_1+\omega x^2_2+\omega^{-1} p^2_2 \right) +2\frac{e^{\Omega_2/T_2}-e^{\Omega_1/T_1}}{\left(e^{\Omega_1/T_1}+1\right)\left(e^{\Omega_2/T_2}+1\right)}\left(\omega x_1x_2+\omega^{-1} p_1p_2 \right) \right).
\end{eqnarray}
In the derivation of~(\ref{Wigner_A}) we used the fact that the average values of momentum and position operators vanish in the long time limit, i.e. $\langle \xi_i(\infty)\rangle=0 $, $i=1,..,4$.  The Wigner function~(\ref{Wigner_A}) corresponding to the case of equal temperatures was obtained in~\cite{xue2013thermal}.

One can easily check that the transformation inverse to~(\ref{Wrho}) is given by
\be\label{rhoW} 
\rho(x'_1,x'_2,x_1,x_2;t)=\int W\left(\frac{x_1+x'_1}{2},p_1,\frac{x_2+x'_2}{2},p_2;t\right) e^{-ip_1(x_1-x'_1)-ip_2(x_2-x'_2)}dp_1dp_2,
\ee
where the density matrix in position representation reads $\rho(x'_1,x'_2,x_1,x_2; t)=\langle x'_1,x'_2| \hat{\rho}(t)| x_1,x_2\rangle$. Inserting~(\ref{Wigner_A}) into~(\ref{rhoW}), we obtain the density matrix for the intermediate equilibrium state
\begin{eqnarray} \label{rhoA_pos}
\rho_A(x'_1,x'_2,x_1,x_2; \infty)=\frac{\omega}{\pi \sqrt{\coth \frac{\Omega_1}{2T_1}\coth \frac{\Omega_2}{2T_2}}} \exp \left[ 
-\frac{\omega}{4}\left(\coth \frac{\Omega_1}{T_1}+\coth \frac{\Omega_2}{T_2}\right)\left(x^2_1+x^2_2+x'^2_1+x'^2_2\right)\right.
\nonumber \\
+\frac{\omega}{2}\left(\frac{1}{\sinh \frac{\Omega_1}{T_1}}+\frac{1}{\sinh \frac{\Omega_2}{T_2}}\right)\left(x_1 x'_1+x_2 x'_2\right)
-\frac{\omega}{2}\left(\coth \frac{\Omega_1}{T_1}-\coth \frac{\Omega_2}{T_2}\right)\left(x_1 x_2+x'_1 x'_2\right)\nonumber \\
+\frac{\omega}{2}\left(\frac{1}{\sinh \frac{\Omega_1}{T_1}}-\frac{1}{\sinh \frac{\Omega_2}{T_2}}\right)\left(x'_1 x_2+x_1 x'_2\right)
\left.\right].
\end{eqnarray}
In the case of equal the temperatures of the baths, i.e. $T_1=T_2\equiv T$, the density matrix~(\ref{rhoA_pos}) is the equilibrium state of a two coupled oscillators 
\be\label{rho_Gibs}
\hat{\rho}_T=\left[\mbox{Tr}\exp{ \left(-\hat{H}_{12}/T \right) }\right]^{-1} \exp \left(-\hat{H}_{12}/T \right).
\ee
with the Hamiltonian 
\be
\hat{H}_{12}=\omega\left(\hat{a}^{\dagger}_1 \hat{a}_1+\hat{a}^{\dagger}_2 \hat{a}_2 \right )+\lambda\left(\hat{a}^{\dagger}_1 \hat{a}_2+\hat{a}^{\dagger}_2 \hat{a}_1 \right )
\ee
embedded in a heat bath characterized by a temperature $T$, i.e. $\rho_A(x'_1,x'_2,x_1,x_2; \infty)=\langle x'_1,x'_2| \hat{\rho}_T| x_1,x_2\rangle$. This equilibrium state is know to be a Gibbsian one. The density matrix~(\ref{rho_Gibs}) of two coupled oscillators in position representation was calculated in~\cite{abdalla1990anisotropic}.
\section{Conclusions}
\pst
To conclude, we summarize the main results of the present paper. We considered, following~\cite{glauber1984damping}, a system of two coupled quantum oscillators (the $A$ oscillator) each embedded in its own thermal bath with different temperatures. Each of the baths are modeled by an infinite set of independent oscillators (the $B$ oscillators). We assumed that the initial state of the $A$ oscillator is a direct product of coherent states, whereas the bath's oscillators are initially in thermal states. The time-dependent state of the $A$ oscillator is a bipartite Gaussian state. We used Simon's separability criterion for two-mode Gaussian states and proved that the state of the $A$ oscillator being separable at initial time $t=0$ remains to be separable for all times $t\geq 0$. As time goes to infinity the state of the two oscillators under consideration  approaches to the intermediate equilibrium state which is a non-Gibbsian and depends on the temperatures of the baths. We gave an explicit expression for the Wigner function as well as the density matrix of this state in position representation. We considered the case of equal temperatures of the baths and showed that the intermediate equilibrium state is Gibbsian in this case.
\bibliography{Citation}

\end{document}